\documentclass[twocolumn]{jpsj2} 
\def\simgt{\lower.5ex\hbox{$\; \buildrel > \over \sim \;$}}

\title{Recent Topics on Very High Energy Gamma-ray Astronomy}

\author{Masaki \textsc{Mori}\thanks{E-mail: morim@icrr.u-tokyo.ac.jp}}

\inst{Institute for Cosmic Ray Research, University of Tokyo\\
5-1-5 Kashiwanoha, Kashiwa, Chiba 277-8582, Japan}

\abst{With the advent of imaging atmospheric Cherenkov telescopes in late 1980's, ground-based observation of TeV gamma-rays came into reality after struggling trials by pioneers for twenty years, and the number of gamma-ray sources detected at TeV energies has increased to be over seventy now. In this review, recent findings from ground-based very-high-energy gamma-ray observations are summarized (as of 2008 March), and up-to-date problems in this research field are presented.}

\kword{gamma-rays, supernova remnants, pulsar nebulae, active galactic nuclei}

\begin{document}
\maketitle

\section{Introduction} 

There has been a dramatic progress in very-high-energy \cite{TeV} 
gamma-ray astronomy in the last decade. With the advent of imaging atmospheric Cherenkov telescopes, the number of gamma-ray sources detected in the TeV energy range
 has increased significantly in the last several years, and now it exceeds seventy, according to the rapporteur talk
by Jim Hinton at the International Cosmic Ray Conference in 2007 \cite{Hin07}. 
These gamma-ray sources form a new class of high-energy objects in the Universe, including active galactic nuclei, radio galaxies, galactic binaries, pulsar wind nebulae, in addition to supernova remnants which are assumed to be the origin of cosmic-rays for a long time. Exploring the emission mechanism from these objects is a big challenge in astrophysics. Non-thermal nature of emission inherently needs multiwavelength observations to study the phenomena, involving astronomers working in other wavelength. Furthermore, many unidentified gamma-ray sources have been reported and are posing new mysteries. 

In this review, highlights of recent findings from ground-based very-high-energy gamma-ray observations are summarized, and up-to-date problems in this research field are presented. 
We concentrate on recent results from Cherenkov
telescopes here, but new results from particle shower arrays 
are also exciting (G.~Sinnis and M.~Takita, in these proceedings.)
Recent review on this field is also found elsewhere \cite{DeA07}.

\section{Imaging Atmospheric Cherenkov Telescopes}

Ground-based imaging Cherenkov telescopes are becoming
a powerful tool to study
very high energy gamma-rays with their capability to discriminate gamma-rays from background hadrons (protons and nuclei) \cite{Wee03}.

Gamma-ray images come from purely electromagnetic showers and
are sharp and oriented toward the object being tracked.
They can be separated from cosmic-ray showers using imaging parameters first defined by A.M.~Hillas \cite{Hil85}: 
{\it width, length, distance, asymmetry}. 
Distributions of theses parameters for gamma-ray showers are different 
from those for hadronic showers, and we can extract
gamma-ray signals statistically from observed images.
The first firm detection of a TeV gamma-ray signal from the Crab nebula
by the Whipple group \cite{Wee89} utilized this {\it imaging technique}.

Stereoscopic observation of Cherenkov images came into practical use
by the HEGRA group \cite{Koh96}. Incoming direction of a gamma-ray 
can be determined by intersection of axes of elongated Cherenkov
images observed by multiple telescopes, separated by about 100\,m, 
which is the size of light pool of Cherenkov light flash,
more precisely compared to single telescope observation. 
Distance of centroid of shower images from the incoming direction 
is a measure of shower maximum height in the atmosphere, 
which also allows better estimation of gamma-ray energy.
In addition, background Cherenkov
flashes caused by local cosmic-ray muons traversing near telescopes
can be effectively rejected requiring coincidence between telescopes.

Table \ref{table:cts} summarizes characteristics of Cherenkov
telescope systems in operation.

\begin{center}
\begin{table*}[tb]
\caption{Current Cherenkov telescope systems}
\label{table:cts}
\begin{tabular}{|l||c|c|c|c|c|c|} \hline
Group &Location & Latitude & Longitude & Height & Telescopes & Start \\ \hline\hline 
SHALON & Russia & $42^\circ$S & $75^\circ$E & 3,338\,m & 11\,m$^2\times$2 & 1992 \\ \hline
TACTIC & India & $25^\circ$S & $73^\circ$E & 1,300\,m & 9.5\,m$^2\times$4 & 2000 \\ \hline
CANGAROO-III & Australia & $31^\circ$S & $137^\circ$E & 160\,m & 57\,m$^2\times$4 & 2004 \\ \hline
H.E.S.S. & Namibia & $23^\circ$S & $16.5^\circ$E & 1,800\,m & 107\,m$^2\times$4 & 2004 \\ \hline
MAGIC & Canary Is. & $29^\circ$N & $18^\circ$W & 2,200\,m & 237\,m$^2\times$1 & 2004 \\ \hline
VERITAS & Arizona & $32^\circ$N & $111^\circ$W & 1,300\,m & 110\,m$^2\times$4 & 2007 \\ \hline
\end{tabular}
\end{table*}
\end{center}

\section{Recent Topics}

Table \ref{table:sum} is a summary of TeV sources
which are fairly established (detected by multiple groups
and/or detected at high significance).
Over 40 sources in this table were found by the H.E.S.S.\ group,
mostly by their Galactic plane survey \cite{Aha05a}.
Note the classification of sources are not unique
and depend on person who makes a list:
identification of TeV sources with objects found
in other wavelength are sometimes difficult, 
because of the limited angular resolution of TeV
observation: 
therefore many TeV sources are left unidentified.
(But there are `dark' sources: see section \ref{sec:unid}.)

\begin{center}
\begin{table}
\caption{Summary of TeV sources as of March 2008 \cite{Mor08}}
\label{table:sum}
\begin{tabular}{|l|r|} \hline 
Source type & Count \\ \hline\hline 
{\it Galactic sources} & \\
\ \ Shell-type supernova remnant & 3 \\
\ \ Supernova remnant & 11 \\
\ \ Pulsar wind nebula & 16 \\
\ \ Binary & 4 \\
\ \ Galactic center & 1 \\
\ \ Galactic ridge & 1 \\
\ \ Young stellar cluster & 1 \\
\ \ Unidentified & 14 \\
{\it Extragalactic sources} & \\
\ \ High-frequency peaked BL Lac (HBL) & 17 \\
\ \ Low-frequency peaked BL Lac (LBL) & 1 \\
\ \ Radio galaxy & 1 \\
\ \ Flat-spectrum radio quasar (FSRQ) & 1 \\ \hline 
{\sl Total} & {\it 71} \\ \hline 
\end{tabular}
\end{table}
\end{center}

\begin{figure*}[tb]
\begin{center}
\includegraphics[height=7.2cm]{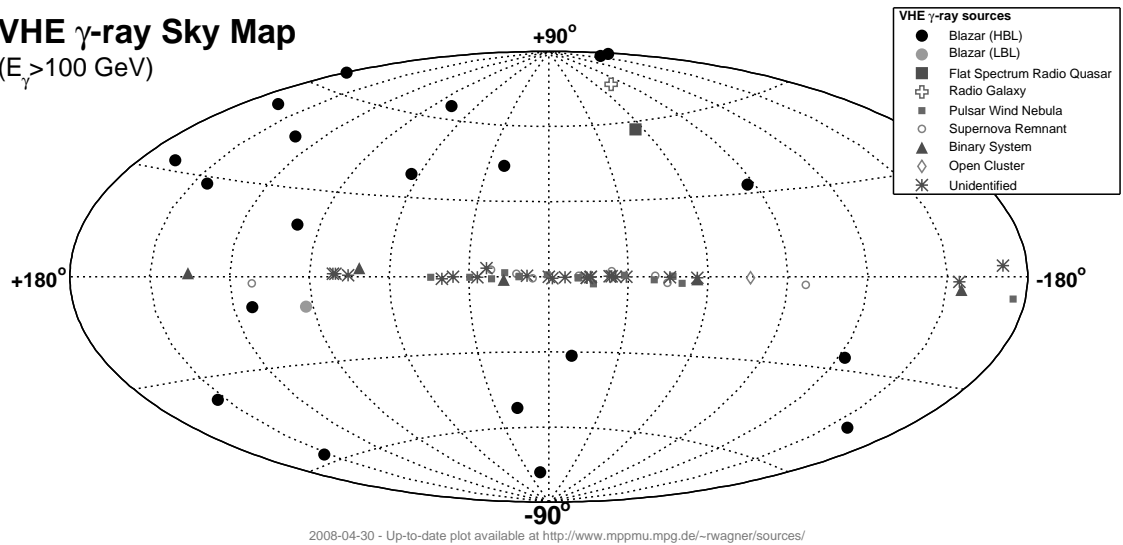}
\end{center}
\caption{TeV skymap (circa early 2008) compiled by R.~Wagner \cite{Wag08}. }
\label{fig:skymap}
\end{figure*}

Fig.~\ref{fig:skymap} is a skymap of TeV sources 
in the Galactic coordinates compiled by R.~Wagner \cite{Wag08}.
Concentration of sources along the Galactic plane is obvious,
but please note the exposure time is far from uniform.

\subsection{Galactic Sources}

\subsubsection{Supernova Remnants}

Needless to say, supernova remnants (SNRs) are long considered to be
primary sources for galactic cosmic rays \cite{Aha04t}.
They are energetic enough to support the cosmic ray luminosity
of the Galaxy and their sizes are large enough to confine
cosmic rays during acceleration by diffusive shockwave.
Bright synchrotron radio emission from many SNRs indicate
existence of non-thermal, high-energy electrons.
Evidences of acceleration of protons in SNR, however, were rare.

The first SNR detected at TeV energies is {\bf RX J1713.7-3946}
(by CANGAROO-I \cite{Mur00}, CANGAROO-II \cite{Eno02}, and
H.E.S.S.\ \cite{Aha04}) where the shell structure was
clearly observed \cite{Aha04}. 
Although the spectrum is better explained by emission from
neutral pions produced in hadronic interaction of protons
\cite{Eno02, Aha04, Aha07d}, inverse Compton emission by electrons
is still a viable process to account for the TeV spectrum. 
{\bf RX J0852.0-4622} (by CANGAROO-II \cite{Kat04} 
and H.E.S.S.\ \cite{Aha05})
and {\bf RCW 86} (by H.E.S.S.\ \cite{Hop07}) 
are other SNRs which show shell-like
emission profile, well correlated with X-ray intensity profiles.
{\bf Cas A}, first claimed by HEGRA \cite{Aha01} and confirmed to be 
a TeV emitter by MAGIC \cite{Alb07b}, and {\bf CTA 37A/37B}, 
recently claimed by H.E.S.S. \cite{Aha08}, could also be shell-like
SNRs, but their small angular sizes do now allow further study.
These shell-like structure are often well correlated with
X-ray images, and one may suppose TeV emission comes from
inverse Compton scattering of low energy photons by
high-energy electrons which radiate synchrotron X-rays \cite{Kat08}.
Still, gamma-rays derived from neutral pion decays 
produced in collision of high-energy protons 
with ambient matter could account for TeV emission \cite{Ber07},
since amplified magnetic fields in SNRs might suppress
inverse Compton emission \cite{Ell07}.

Uchiyama {\it et al.} analyzed the X-ray images of
RX J1713.7-3946 observed by Chandra in multiple years,
and found there are `hot spots' showing time variation in
one-year scale \cite{Uch07}. 
This time scale, compared with the cooling
time of high energy electrons, means the ambient magnetic field
is stronger than 1~mG. This magnetic field suppresses
the inverse Compton emission in the TeV region, and
may indicate TeV emission is produced by hadronic process.
However, Butt {\it et al.} argue that the existing multiwavelength 
data on this object do not support such conclusion \cite{But08}.

On the other hand, TeV emission profiles of {\bf IC443} 
(by MAGIC \cite{Alb07} and VERITAS \cite{Hum07})
and {\bf W28} (by H.E.S.S.\ \cite{Aha08a}) 
do not show shell-like structure
and seems to correlate with molecular cloud profiles
observed by a CO emission line. This interaction may
indicate the hadronic origin of gamma-rays produced in
collisions of high-energy protons with molecular cloud as targets.

{\bf HESS J0632+057} was found close to the rim of the Monoceros
Loop SNR/Rosette Nebula region \cite{Aha07c}. 
It is point-like and has no clear counterpart 
at other wavelengths, but is possibly associated
with the GeV source 3EG J0634+0521, a weak X-ray source 
1RXS J063258.3+054857 and the Be-star MWC 148.

Thus, although supernova remnants are proved to be the sites producing
high energy particles in the Universe, species of accelerated particles
are still unclear \cite{Tat08}: solving the long-standing mystery of
cosmic ray origin is still an important issue of gamma-ray astronomy.

\subsubsection{Pulsar Wind Nebulae}

\begin{center}
\begin{table}
\caption{TeV sources associated with pulsar wind nebulae 
\cite{Hin07, Mor08}.
These associations were established through a range of methods, which are given in the table in abbreviated
form: {\it Pos}: The position of the centroid of the VHE emission 
can be established with sufficient precision
that there is no ambiguity as to the low energy counterpart. 
{\it Mor}: There is a match between the 
gamma-ray morphology and that seen at other (usually
X-ray) wavelengths. 
{\it EDMor}: Energy-dependent morphology which approaches the position/morphology seen at other
wavelengths at some limit, and is consistent with our physical understanding of the source.}
\label{table:pwn}
\begin{tabular}{|l|l|c|l|} \hline 
Object & Pulsar & Method & Discovered \\ \hline\hline 
Crab nebula & B0531+21 & Pos & Whipple, 1989 \\
MSH 15$-$52 & B1509$-$58 & Mor & HESS, 2005 \\
Vela X & B0833$-$45 & Mor & HESS, 2006 \\
HESS J1825$-$137 & B1823$-$13 & EDMor & HESS, 2005 \\
PSR J1420-6049 & J1420$-$6048 & Mor & HESS, 2006 \\
The Rabbit & J1420$-$6048 & Mor & HESS, 2006 \\
G0.9+0.1 & -- & Pos & HESS, 2005 \\ \hline 
\end{tabular}
\end{table}
\end{center}

Table \ref{table:pwn} is a list of TeV sources
with well established association with pulsar wind nebulae.
Including associations with weaker evidence,
18 out of 71 sources detected at TeV energies 
seem to be associated
with pulsar wind nebulae \cite{Hin07}, which was a rather
surprising discovery revealed by the H.E.S.S.\ Galactic 
survey \cite{Aha05a}. Most of them are associated with
relatively young ($<10^5$ years) and large spin-down
pulsars \cite{Fun07}, which means the gamma-ray luminosity is
supported by the pulsar spin-down energy. Their
profiles show extended (order of 10~pc) structure,
often displaced from pulsar positions. No pulsation
has been reported in the TeV region, even
for the Crab nebula which is a strong TeV source
and best studied by a number of groups.
Thus the TeV emission are naturally ascribed to inverse
Compton emission by high-energy electrons
accelerated in the vicinity of pulsars.

This hypothesis on emission mechanism is further supported
by the energy-dependent morphology in {\bf HESS J1825-137}
\cite{Aha06}. The photon indices from a power-law fit 
in different regions show a softening of 
the spectrum with increasing distance from the pulsar. 
The observed energy dependent morphology
may be an evidence for cooling of electrons
in the nebula.

{\bf HESS J1837$-$069}, which was classified as unidentified
before, has now been added to this category
with the discovery of an energetic 70.5~ms pulsar in 
AX J1838.0$-$0655 using X-ray data by RXTE \cite{Got08}.

\subsubsection{Gamma-ray Binaries}

\paragraph{PSR B1259$-$63/SS2883}
This is a 48~ms pulsar and a Be star binary  
in a highly eccentric orbit. Gamma-ray flux
at detectable level was predicted when the 
binary is near the periastron passage via interaction of pulsar wind
with the radiative environment of the binary system 
\cite{Kir99}. H.E.S.S.\ detected such a variable
flux around the 2004 periastron \cite{Aha05c}, with
a double-peak light curve as predicted by some models \cite{Kaw04}.

\paragraph{LS5039}
This is a high mass X-ray binary comprising a massive star 
and a compact object, and is resolved 
into a bipolar radio outflow emanating from a central
core, thus it is often classified as the microquasar
class.
H.E.S.S.\ detected a modulated gamma-ray signal with the 3.9 day
orbital period in 2005 \cite{Aha06a}.
The emission maximum ($\phi \sim 0.7$) appears
to lag behind the apastron epoch and to align better
with the inferior conjunction ($\phi= 0.716$), when
the compact object lies in front of the massive star. 
The flux minimum occurs at
phase ($\phi \sim 0.2$), slightly further along the orbit
than superior conjunction ($\phi = 0.058$).
This behavior is not easy to explain by a simple
assumption and offers a challenge to model builders.

\paragraph{LSI+61 303}
MAGIC observed this microquasar,
known in the GeV region as 2CG 135+01 (by COS~B) 
or 3EG J0241+6103 (by EGRET),
for six orbital cycles
($P_{\rm orb}=26.5$ days) and obtained a modulated
gamma-ray signal which is significant only between
orbital phase 0.4 and 0.7 \cite{Alb06}.
The flux maximum is detected at phases 0.5--0.6, 
overlapping with an X-ray outburst and the onset 
of a radio outburst,
is shifted from the phase when the two stars are 
closest to one another, implying a strong orbital 
modulation of the emission or the absorption processes.
The VERITAS group confirmed the modulated emission \cite{Mai07}.

\paragraph{Cyg X-1}
This well-known black hole X-ray binary was
observed by MAGIC for a total of 40 hours during 26
nights in 2006. Although the steady TeV emission
was not observed, a $4\sigma$-level ($3.2\sigma$ after
trial correction) evidence was obtained for one night
on September 24, coinciding with an X-ray flare
\cite{Alb07a}. The significance of detection
may not be strong enough to claim Cyg X-1 is a
TeV gamma-ray source \cite{Hin07}.

\subsubsection{Stellar Cluster}

The H.E.S.S.\ group reported {\bf Westerlund 2}, a young
open stellar cluster which contains dozen O-stars
and two Wolf-Rayet stars, is an extended gamma-ray
source, HESS J1023-575 \cite{Aha07a}. This is a new source class
and the gamma-ray emission mechanism is a new
challenge to theorists: extended profile suggests
there might be collective effects, possibly of
stellar winds in the cluster.

\subsubsection{Galactic Center}

The {\bf Galactic Center} is a confirmed TeV source
(by CANGAROO-II \cite{Tsu04}, Whipple \cite{Kos04}, and
H.E.S.S.\ \cite{Aha04a}), but the nature of
the emission is not known. 
The energy spectrum shows a simple power-law
shape, not compatible with dark matter annihilation
signal \cite{Aha06b}, but the position
coincides with the central radio source, Sgr A*
\cite{Eld07}.

Moreover, when the Galactic center source and 
the nearby source, G0.9+0.1, are subtracted from
the H.E.S.S.\ data, emission extended in galactic longitude
for roughly $2^\circ$
and also in galactic latitude with a characteristic
width of about $0.2^\circ$ were seen \cite{Aha06c}.
The power-law index of the spectrum (2.3) of
this Galactic ridge gamma-ray emission is
harder than the local cosmic-ray spectrum (2.7): this
may indicate the propagation effects are less
pronounced than in the Galaxy as a whole due to
the proximity of particle accelerators.

\subsubsection{Unidentified Sources}
\label{sec:unid}

Significant number of the TeV sources distributed
near the Galactic plane are unidentified \cite{Aha05a}. 
Some have no compelling counterparts, but some are completely
dark in other wavelength \cite{Aha08b}. 
Former sources could be 
identified if the angular resolution of Cherenkov telescopes
are improved in future detectors, but latter sources
bring up a new mystery in astrophysics.
Multiwavelength observations of these sources are going on
to reveal their identities.

{\bf HESS J1908+063} was discovered during 
the extended H.E.S.S.\ survey of the Galactic plane, 
and it coincides with the recently reported MILAGRO 
unidentified source MGRO J1908+06 \cite{Dja07}.

{\bf TeV J2032+4130} was discovered by HEGRA \cite{Aha02}
in the Cygnus complex region from their Galactic plane survey,
and confirmed by Whipple \cite{Lan04} and MAGIC \cite{Alb08}.
It is also within the extended MILAGRO source MGRO J2031+41 \cite{Abd07}.

\begin{figure}[tb]
\begin{center}
\includegraphics[height=6cm]{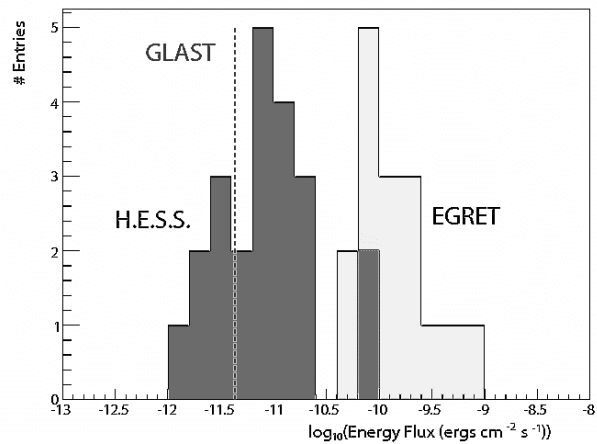}
\end{center}
\caption{Distribution of integrated energy flux
for sources in the Inner Galaxy \cite{Fun07a}. 
For EGRET the energy flux between 1 GeV and 10 GeV, for the H.E.S.S. sources, the
energy flux between 1 TeV and 10 TeV is shown. Also shown is the sensitivity prediction
for the GLAST-LAT for a typical location in the Inner Galaxy 
($\ell=10^\circ, b=0^\circ$).}
\label{fig:Fun07a}
\end{figure}

Funk {\it et al.} studied the TeV-GeV connection in
Galactic sources comparing the TeV sources
with the EGRET catalog \cite{Fun07a}.
Surprisingly, few common sources are found in terms
of positional coincidence and spectral consistency.
Distribution of integrated energy flux (Fig.\ \ref{fig:Fun07a}) shows
almost separate peaks for TeV sources and EGRET sources,
which means the current TeV sensitivity is much better than
GeV sensitivity.
The TeV upper limits put strong constraints on simple power-law
extrapolation of several of the EGRET spectra 
and thus strongly suggest cutoffs in the unexplored 
energy range from 10 GeV to 100 GeV.

\subsection{Extragalactic Sources}

Table \ref{table:extrag} is a list of known
extragalactic sources in the TeV region
in the order of redshift ($z$).

\begin{center}
\begin{table}
\caption{Known extragalactic TeV sources as of March 2008 \cite{Mor08}.
(See Table \ref{table:sum} for abbreviation of source class.)}
\label{table:extrag}
\begin{tabular}{|l|c|l|l|} \hline 
Object & $z$ & Class & Discovered \\ \hline\hline 
M87 & 0.004 & Radio & HEGRA, 2003 \\
Mrk 421 & 0.031 & HBL & Whipple, 1992 \\
Mrk 501 & 0.034 & HBL & Whipple, 1996 \\
1ES 2344$+$514 & 0.044 & HBL & Whipple, 1998 \\
Mrk 180 & 0.046 & HBL & MAGIC, 2006 \\
1ES 1959$+$650 & 0.047 & HBL & 7TA, 1999 \\
BL Lac & 0.069 & LBL & MAGIC, 2007 \\
PKS 0548$-$322 & 0.069 & HBL & HESS, 2007 \\
PKS 2005$-$489 & 0.071 & HBL & HESS, 2005 \\
PKS 2155$-$304 & 0.116 & HBL & Durham, 1999 \\
H 1426$+$428 & 0.129 & HBL & Whipple, 2002 \\
1ES 0229$+$428 & 0.140 & HBL & HESS, 2007 \\
H 2356$-$309 & 0.165 & HBL & HESS, 2006 \\
1ES 1218$+$304 & 0.182 & HBL & MAGIC, 2006 \\
1ES 1101$-$232 & 0.186 & HBL & HESS, 2006 \\
1ES 0347$-$121 & 0.188 & HBL & HESS, 2007 \\
1ES 1011$+$496 & 0.212 & HBL & MAGIC, 2007 \\
PG 1553$+$113 & $>0.25$ & HBL & HESS, 2006 \\
3C279 & 0.536 & FSRQ & MAGIC, 2007 \\ \hline
\end{tabular}
\end{table}
\end{center}

\subsubsection{Blazars}

Most of sources listed in Table \ref{table:extrag}
are blazar-type active galactic nuclei (AGN),
including the second and third established TeV sources,
Mrk 421 and Mrk 501.
The prominent feature of gamma-ray emission from this class is
the rapid time variability, which should be related
to the central engine of activity, accretion process
around supermassive black holes. In many cases, 
gamma-ray emission is accounted by inverse Compton process
by high-energy electrons, which naturally explains X-ray
and TeV correlated time variation, but hadronic process
is still a viable option \cite{Aha04t}.

In July 2006, the H.E.S.S.\ group reported 
an exceptionally strong flare of {\bf PKS 2155$-$304} \cite{Ben06}.
The peak flux was seven times the flux observed from the Crab
nebula, and the flux varied at time scale as short as one
minute \cite{Aha07b}. The CANGAROO-III group confirmed
this flaring activity but in a different time zone \cite{Sak08}.
This short time variability set limit on the size and 
Doppler factor of gamma-ray emission region 
near the central massive black hole of the blazar.
Assuming the emission region has a size comparable to the Schwarzschild radius of a $\sim 10^9 M_\odot$ black hole, Doppler factors greater than 100 are required to accommodate the observed variability time scales
 \cite{Aha07b}.

The MAGIC group reported a detection of {\bf 3C279} \cite{Tes07},
a prominent flat-spectrum-radio-quasar-type AGN in the GeV region
observed by EGRET \cite{Har01}. It was believed
that its distance ($z=0.536$) was too far to be detected in
very-high-energy gamma-rays due to the absorption effect
by the extragalactic infrared background radiation
(see next subsection), and we should draw attention
to further observations of this object, since the
current statistics is too limited to draw spectral information.

\subsubsection{Extragalactic background radiation}

Stecker {\it et al.} \cite{Ste92} pointed out that
extragalactic diffuse infrared background radiation
blocks the propagation of TeV gamma-ray
over large distances ($z\simgt0.1$ or $d\simgt100$ Mpc)
by producing electron-positron pairs.
However, measuring the intensity of 
the background infrared radiation is a difficult
task, since subtracting foreground sources from observed data
is a very complicated process.
Recent progress has been summarized by
many authors: see ref.~\cite{Rau08},
for example.
In turn, we can infer the intensity from
observation of TeV spectra of distant sources,
assuming the spectra at sources are known:
this cannot be known, in fact, so what we
can do in practice is to place an upper limit of the intensity
of infrared background radiation.
Such studies (ref.~\cite{Aha06d}, for example)
indicate the infrared background radiation
might be at lower level than anticipated.
This is a result placing constraint on galaxy 
formation theory, and is a good example of 
potential power of TeV gamma-ray astronomy toward cosmology
\cite{Ste07}.

\begin{figure}[tb]
\begin{center}
\includegraphics[height=6cm]{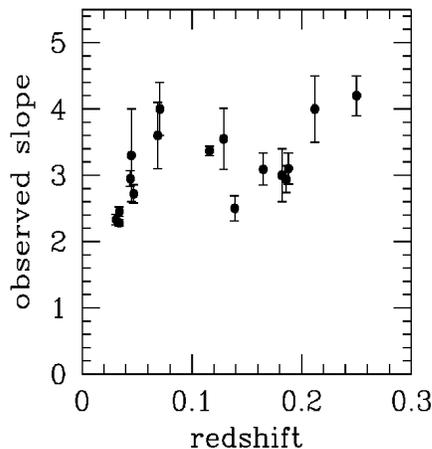}
\end{center}
\caption{Observed spectral indices for AGN in the VHE region \cite{DeA07}.
}
\label{fig:sindex}
\end{figure}

Fig.\ref{fig:sindex} is a plot of spectral indices
of AGN in the TeV energy region against redshift
\cite{DeA07}. If the intrinsic spectra of AGN
are similar, there should be a rising tendency
in this plot as redshift increases, but we cannot
draw clear conclusion from this plot: surely we need more
samples.

\subsubsection{M87}

HEGRA reported a detection of {\bf M87}, a nearby radio galaxy
in the Virgo cluster ($z=0.00436$ or $\sim 16$~Mpc) in 2003 \cite{Aha03}.
H.E.S.S.\ confirmed the gamma-ray emission of this source,
which showed the short-time variability of 2-day scale 
based on observation between 2003 and 2006 \cite{Aha06e}.
This time scale is the order of the light crossing time of the central black hole.
The VERITAS group also detected a gamma-ray signal in 2007 \cite{Col07}.

\subsubsection{Centaurus A}

Recent data from the Pierre Auger Observatory suggests this 
nearby giant radio galaxy ($z=0.0018$) could be 
a source of ultra-high-energy cosmic rays \cite{PA07}. 
If true, one may expect it to emit gamma-rays \cite{Gup08}.
However, in the TeV region, only upper limits
on gamma-ray flux are reported except one evidence
reported in 1970's \cite{Gri75, Aha05b, Kab07}.

\subsubsection{Clusters of Galaxies}

As the largest systems in the Universe, clusters of galaxies can
harbor high-energy cosmic-rays for cosmological time which
may be accelerated in merger shocks and/or accretion shocks,
and can emit gamma-rays at detectable level via
various possible processes \cite{Ino05}.

The Whipple group observed {\bf Perseus} ($z=0.018$)
 and {\bf Abell 2029} ($z=0.078$),
but extended gamma-ray emission signals were not observed \cite{Per06}.
The H.E.S.S.\ group looked for gamma-ray emission 
from two nearby clusters, {\bf Abell 496} ($z=0.033$) and {\bf Coma}
($z=0.024$) \cite{Dom07}. The CANGAROO-III group 
selected {\bf Abell 4038} ($z=0.028$) and {\bf Abell 3667} ($z=0.055$)
\cite{Kiu07} for target.
Again, both groups could not see a hint of signal
of gamma-ray emission.

\section{Future Projects}

\begin{figure}[tb]
\begin{center}
\includegraphics[height=6cm]{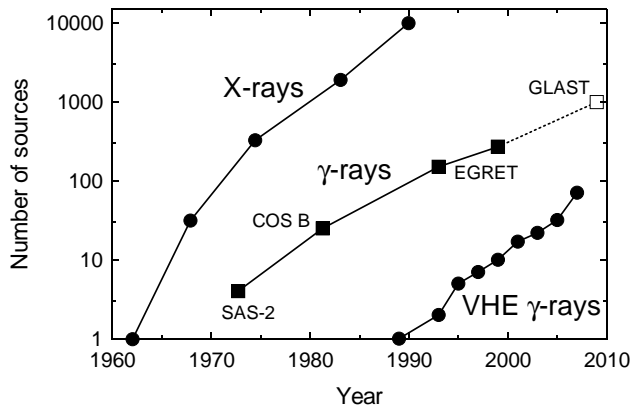}
\end{center}
\caption{Yearly increase of X-ray, gamma-ray and very-high-energy
(VHE) gamma-ray sources. (This type of plot is called
as {\it Kifune plot}, named after Prof.\ Tadashi Kifune who invented it.)}
\label{fig:kifune}
\end{figure}

With these impressive discoveries, next-generation large-scale
atmospheric Cherenkov telescope systems are
under hot discussion in recent years
to enhance the sensitivity in TeV gamma-rays furthermore. 
The H.E.S.S.\ group is constructing a huge 28~m diameter
telescope, called H.E.S.S.~II, at the center of the present
array \cite{Vin05}, and the MAGIC II telescope, an advanced copy
of the present MAGIC 17~m telescope 80~m apart, 
will be completed this year
in order to enjoy the merit of stereoscopic observation \cite{Goe07}.
In the longer range, the size of next-generation
projects naturally requires international collaboration.
CTA (Cherenkov Telescope Array) \cite{CTA} is one of such projects
and is lead mainly by H.E.S.S.\ and MAGIC collaborators.
AGIS (Advanced Gamma-ray Imaging System) \cite{AGIS} is another
initiative proposed mainly by VERITAS collaborators.
These projects aim to increase the sensitivity by a
factor of ten, as well as to enhance the gamma-ray
energy range both lower and higher regions.
The trend seen in the {\it Kifune plot} (Fig.\ref{fig:kifune})
predicts a bright future of TeV astronomy with an order of
one thousand sources in mid 2010's.

\section{Summary}

Now the very-high-energy window of the Universe is open, and
Cherenkov telescopes are providing additional 2--3 decades
to the photon spectra of high-energy objects in the sky.
We can see wider variety of source classes
among seventy TeV sources than anticipated, 
indicating cosmic accelerators are {\it ubiquitous}
in the Universe. However, understanding the underlying physical
processes is still at a preliminary stage.
Together with GLAST (P.~Michelson, in these proceedings.), to be launched
in 2008 to observe GeV gamma-rays from space, 
we can expect many findings from the exploration of high-energy Universe
with next-generation ground-based gamma-ray detectors.

\section*{Acknowledgment}
This work is partly supported by Grant-in-Aid for Scientific Research from the Ministry of Education, Culture, Sports,
Science and Technology.


\begin{thebibliography}{99} 

\bibitem{TeV} In this article, {\it very high energy} (VHE)
means the energies from
100~GeV to 10~TeV, and has the same meaning of {\it TeV region}.
\bibitem{Hin07} J. Hinton: Rapporteur talk at 30th Int. Cosmic Ray Conf., Merida, Mexico, 2007; arXiv:0712.3352 (2007).
\bibitem{DeA07} A.~De Angelis O.~Mansutti, and M.~Persic: La Rivista del Nuovo Cimento {\bf 31} (2008) 187--246; arXiv:0712.0315 (2007).
\bibitem{Wee03} T.C. Weekes: {\it Very High Energy Gamma-ray Astronomy}  (Institute of Physics Publishing, Bristol and Philadelphia, 2003).
\bibitem{Hil85} A. M. Hillas: {\it Proc. 19th Int. Cosmic Ray Conf.} (La Jolla, USA, 1985) Vol.3, p.445.
\bibitem{Wee89} T.C. Weekes {\it et al.}: Astrophys. J. {\bf342} (1989) 379.
\bibitem{Koh96} A. Kohnle {\it et al.}: Astroparticle Phys. \textbf{5} (1996) 119.
\bibitem{Aha05a} F. Aharonian {\it et al.}: Science {\bf 309} (2005) 746--749; Astrophys. J. {\bf 636} (2006) 777.
\bibitem{Mor08} See http://tevcat.uchicago.edu/ or http://www.icrr.u-tokyo.ac.jp/{\~{}}morim/TeV-catalog/index.html for more information.
\bibitem{Wag08} http://www.mppmu.mpg.de/{\~{}}rwagner/sources/
\bibitem{Aha04t} F.A.~Aharonian: {\it Very High Energy Cosmic Gamma Radiation -- A Crucial Window on the Extreme Universe} 
(World Scientific Publishing, Singapore, 2004).
\bibitem{Mur00} H. Muraishi {\it et al.}: Astron. Astrophys., {\bf 354}, (2000) L57.
\bibitem{Eno02} R. Enomoto {\it et al.}: Nature {\bf 416} (2002) 823.
\bibitem{Aha04} F. Aharonian {\it et al.}: Nature {\bf 432} (2004) 75.
\bibitem{Aha07d} F.A.~Aharonian {\it et al.}: Astron. Astrophys. {\bf 464} (2007) 235.
\bibitem{Kat04} H. Katagiri {\it et al.}: Astrophys. J. {\bf 619} (2005) L163.
\bibitem{Aha05} F. Aharonian {\it et al.}: Astron. Astrophys. {\bf 437} (2005) L7.
\bibitem{Hop07} S. Hoppe {\it et al.}: arXiv:0709.4103 (2007).
\bibitem{Aha01} F. Aharonian {\it et al.}: Astron. Astrophys. {\bf 370} (2001) 112.
\bibitem{Alb07b} J. Albert {\it et al.}: Astron. Astrophys. {\bf 474} (2007) 937.
\bibitem{Aha08} F. Aharonian {\it et al.}: arXiv:0803.0702 (2008).
\bibitem{Kat08} B.~Katz and E.~Waxman: JCAP 01 (2008) 018 and references therein.
\bibitem{Ber07} E.G.~Berezhko and H.J.~V\"{o}lk: arXiv:0707.4647 (2007). 
\bibitem{Ell07} D.~Ellison, D.~J. Patnaude, P.~Slane, P.~Blasi, and S.~Gabici: Astrophys. J. {\bf 661} (2007) 879.
\bibitem{Uch07} Y.~Uchiyama, F.A.~Aharonian, T.~Tanaka, T.~ Takahashi, and Y.~Maeda: Nature {\bf 449} (2007) 576.
\bibitem{But08} Y.~Butt, T.~Porter, B.~Katz, and E.~Waxman: arXiv:0801.4954 (2007).
\bibitem{Alb07} J.~Albert {\it et al.}: Astrophys. J. {\bf 664} (2007) L87.
\bibitem{Hum07} T.B.~Humensky {\it et al.}: arXiv:0709.4298 (2007).
\bibitem{Aha08a} F. Aharonian {\it et al.}: arXiv:0801.3555 (2008).
\bibitem{Aha07c} F.~Aharonian {\it et al.}: Astron. Astrophys. {\bf 469} (2007) L1.
\bibitem{Tat08} See, for example, V.~Tatischeff: arXiv:0804.1004 (2008) for a recent review.
\bibitem{Fun07} GLAT Collaboration: S.~Funk: arXiv:0709.3127 (2007).
\bibitem{Aha06} F. Aharonian {\it et al.}: Astron. Astrophys. {\bf 460} (2006) 365.
\bibitem{Got08} E.V.~Gotthelf, J.P.~Halpern, F.~Camilo, C.~Markwardt, and J.~Swank: The Astronomer's Telegram \#1392 (2008); L.~Kuiper, W.~Hermsen, M.~Klein-Walt, and R.~Wijands: The Astronomer's Telegram \#1405 (2008).
\bibitem{Kir99} J.G.~Kirk, L.~Ball and O.~Skjaeraasen: Astropart. Phys. {\bf 10} (1999) 31.
\bibitem{Aha05c} F.~Aharonian {\it et al.}: Astron. Astrophys. {\bf 442} (2005) 1.
\bibitem{Kaw04} A.~Kawachi {\it et al.}: Astrophys. J. {\bf 607} (2004) 949.
\bibitem{Aha06a} F. Aharonian {\it et al.}: Astron. Astrophys. {\bf 460},(2006) 743.
\bibitem{Alb06} J.~Albert {\it et al.}: Science {\bf 312} (2006) 1771. 
\bibitem{Mai07} G.~Maier {\it et al.}: arXiv:0709.3661 (2007).
\bibitem{Alb07a} J.~Albert {\it et al.}: Astrophys. J. {\bf 665} (2007) L51.
\bibitem{Aha07a} F.~Aharonian {\it et al.}: Astron. Astrophys. {\bf 467}, (2007) 1075.
\bibitem{Tsu04} K.~Tsuchiya {\it et al.}: Astrophys. J. {\bf 606} (2004) L115.
\bibitem{Kos04} K.~Kosack {\it et al.}: Astropys. J. {\bf 608} (2004)L97.
\bibitem{Aha04a} F.~Aharonian {\it et al.}: Astron. Astrophys. {\bf 425} (2004) L13.
\bibitem{Aha06b} F.~Aharonian {\it et al.}: Phys. Rev. Lett. {\bf 97} (2006) 221102.
\bibitem{Eld07} C.~van Eldik {\it et al.}: arXiv:0709.3729 (2007). 
\bibitem{Aha06c} F.~Aharonian {\it et al.}: Nature {\bf 439} (2006) 695.
\bibitem{Aha08b} F.~Aharonian {\it et al.}: Astron. Astrophys. {\bf 477} (2008) 353.
\bibitem{Dja07} A.~Djannati-Atai {\it et al.}: arXiv:0710.2418 (2007).
\bibitem{Aha02} F.A.~Aharonian {\it et al.}: Astron. Astrophys. {\bf 393} (2002) L37.
\bibitem{Lan04} M.J.~Lang {\it et al.}: Astron. Astrophys. {\bf 423} (2004) 415; A.~Konopelko {\it et al.}: Astrophys. J. {\bf 658} (2007) 1062.
\bibitem{Alb08} J.~Albert {\it et al.}: Astrophys. J. {\bf 675} (2008) L25.
\bibitem{Abd07} A.A.~Abdo {\it et al.}: Astrophys. J. {\bf 664} (2007) L91.
\bibitem{Fun07a} S.~Funk, O.~Reimer, D.F.~Torres and J.A.~Hinton: arXiv:0710.1584 (2007).
\bibitem{Ben06} W.~Benbow, L.~Costamante and B.~Giebels: The Astronomer's Telegram \#867, July 27, 2006.
\bibitem{Aha07b} F.~Aharonian {\it et al.}: Astrophys. J. {\bf 664} (2007) L71.
\bibitem{Sak08} Y.~Sakamoto {\it et al.}: Astrophys. J. {\bf 676} (2008) 113.
\bibitem{Tes07} M.~Teshima {\it et al.}: arXiv:0709.1475 (2007).
\bibitem{Har01} R.C.~Hartman {\it et al.}: Astrophys. J. {\bf 558} (2001) 583 and references therein.
\bibitem{Ste92} F.W.~Stecker, O.C.~de Jager, and M.H.~Salamon: Astrophys. J. {\bf 390} (1992) L49.
\bibitem{Rau08} M.~Raue and D.~Mazin: arXiv:0802/0129 (2008).
\bibitem{Aha06d} F.~Aharonian {\it et al.}: Nature {\bf 440} (2006) 1018.
\bibitem{Ste07} See, for example, F.W.~Stecker: arXiv:0709.0904 (2007).
\bibitem{Aha03} F.~Aharonian {\it et al.}: Astron. Astrophys. {\bf 403} (2003) L1.
\bibitem{Aha06e} F.~Aharonian {\it et al.}: Science {\bf 314} (2006) 1424.
\bibitem{Col07} P.~Colin {\it et al}: arXiv:0709.3663 (2007).
\bibitem{PA07} The Pierre Auger Collaboration: Science {\bf 318}  (2007) 938.
\bibitem{Gup08} N.~Gupta: arXiv:0804.3017 (2008); M.~Kachelriess, S. Ostapchenko, and R. Tomas: arXiv:0805.2608 (2008)
\bibitem{Gri75} J.E.~Grindlay, H.F.~Helmken, R.H.~Brown, J.~Davis, and L.R.~Allen: Astrophys. J. {\bf 197} (1975) L9.
\bibitem{Aha05b} F.~Aharonian {\it et al.}: Astron. Astrophys. {\bf 441}  (2005) 465.
\bibitem{Kab07} S.~Kabuki {\it et al.}: Astrophys. J. {\bf 668} (2007) 968 and references therein.
\bibitem{Ino05} S.~Inoue, F.A.~Aharonian and N.~Sugiyama: Astrophys. J. {\bf 628} (2005) L9 and references therein.
\bibitem{Per06} J.S.~Perkns {\it et al.}: Astrophys. J. {\bf 644} (2006) 148.
\bibitem{Dom07} W.~Domainko {\it et al.}: arXiv:0710.4057 (2007).
\bibitem{Kiu07} R.~Kiuchi {\it et al.}: {\it Proc. 30th Int. Cosmic Ray Conf.} (Merida, Mexico, 2007), paper 166.
\bibitem{Vin05} P.~Vincent {\it et al.}: {\it Proc. 29th Int. Cosmic Ray Conf.} (Pune, India, 2005), Vol.~5, 163.
\bibitem{Goe07} F.~Goebel {\it et al.}: arXiv:0709.2605 (2007).
\bibitem{CTA} http://www.cta-observatory.org/
\bibitem{AGIS} http://gamma1.astro.ucla.edu/agis/index.php/Main\_Page

\end{thebibliography}
\end{document}